\documentclass[preprint,preprintnumbers,amsmath,amssymb,superscriptaddress]{revtex4}
\usepackage{txfonts}
\usepackage{graphicx}
\usepackage{dcolumn}
\usepackage{bm}
\usepackage{array}
\usepackage{changes}
\usepackage{verbatim}
\usepackage{upgreek}
\usepackage{float}
\usepackage{hyphenat}
\usepackage{ragged2e}

\usepackage[colorlinks=true,plainpages=false,linkcolor=blue,urlcolor=blue,citecolor=blue,pdfpagemode=UseNone,pdfstartview=FitBH]{hyperref}

\begin{document}

	\title{Microscopic evidence for Fulde-Ferrel-Larkin-Ovchinnikov state and multiband effects in KFe$_2$As$_2$}

	\author{X. Y. Liu}
	\affiliation{Institute of Physics, Chinese Academy of Sciences,\\
		and Beijing National Laboratory for Condensed Matter Physics, Beijing 100190, China}
	\affiliation{School of Physical Sciences, University of Chinese Academy of Sciences, Beijing 100190, China}
	
	\author{Z. Kao}
	\affiliation{State Key Laboratory of Surface Physics and Department of Physics, Fudan University, Shanghai, China.}

	\author{J. Luo}
	\affiliation{Institute of Physics, Chinese Academy of Sciences,\\
		and Beijing National Laboratory for Condensed Matter Physics, Beijing 100190, China}
	
	\author{J. Yang}
	\affiliation{Institute of Physics, Chinese Academy of Sciences,\\
		and Beijing National Laboratory for Condensed Matter Physics, Beijing 100190, China}
	
	\author{A. F. Fang}
	\affiliation{School of Physics and Astronomy, Beijing Normal University, Beijing 100875, China}
     \affiliation{Key Laboratory of Multiscale Spin Physics, Ministry of Education, Beijing Normal University, Beijing 100875, China}
	
	\author{J. Zhao}
	\affiliation{State Key Laboratory of Surface Physics and Department of Physics, Fudan University, Shanghai, China.}
	\affiliation{Shanghai Research Center for Quantum Sciences, Shanghai, China.}
	\affiliation{Institute of Nanoelectronics and Quantum Computing, Fudan University, Shanghai, China.}
	
	\author{R. Zhou}
	\email{rzhou@iphy.ac.cn}
	\affiliation{Institute of Physics, Chinese Academy of Sciences,\\
		and Beijing National Laboratory for Condensed Matter Physics, Beijing 100190, China}
	\affiliation{School of Physical Sciences, University of Chinese Academy of Sciences, Beijing 100190, China}

	\author{Guo-qing Zheng}
	\affiliation{Department of Physics, Okayama University, Okayama 700-8530, Japan}
	
	\date{\today}
	
	\begin{abstract}
		The Fulde-Ferrell-Larkin-Ovchinnikov (FFLO) state is a superconducting phase characterized by broken translational-symmetry, where Cooper pairs form with non-zero momentum between Zeeman-split Fermi surfaces. This state is highly sensitive to band structure and pairing symmetry. In multiband superconductors, the FFLO state can significantly deviate from its standard form, but experimental verification has remained challenging. Here, we present $^{75}$As nuclear magnetic resonance (NMR) measurements on the multiband superconductor KFe$_2$As$_2$. In the low-temperature, high-magnetic-field region above the upper critical field $B_{c2}$, we observe a clear increase in the second moment of the NMR spectrum, along with a strong enhancement in the spin-lattice relaxation rate divided by temperature 1/$T_1$$T$. These results indicate an emergence of superconducting spin smecticity and Andreev bound states from the spatially modulation of the superconducting gap, providing microscopic evidence for the FFLO state. The obtained phase diagram reveals a distinct boundary line between the FFLO and homogenous superconducting (HSC) states with  a low critical temperature of the FFLO state $T^\ast \approx 0.2 T_c$, which can be attributed to the multiband effects in KFe$_2$As$_2$. Our results show that the iron-based superconductors are a good material platform for studying the FFLO state and highlight the importance of the  multiband effects on this exotic phase.
	\end{abstract}
	
	\maketitle
	

	
As the external magnetic field approaches the Pauli limit field $B_{\rm Pauli}$, the conventional superconducting (SC) state becomes metastable, giving way to a new superconducting state with spontaneous translational-symmetry breaking, which is known as the FFLO state \cite{25,26,5,6,Zwicknagl2010}, where Cooper pairs are formed with electrons from two Zeeman-split Fermi surfaces. Consequently, these pairs have a non-zero center-of-mass momentum, resulting in a spatially modulated energy gap, a defining feature that distinguishes the FFLO state from other unconventional SC states. Despite theoretical predictions of rich physical phenomena in the FFLO state, direct experimental observation has been challenging. The FFLO candidates should have a sufficiently large Maki parameter, $\alpha_{M} = \sqrt{2}  B_{\rm orb}\left ( 0 \right ) /B_{\rm Pauli}\left ( 0 \right ) > 1.8$, which is a measure of the paramagnetic pair-breaking strength\cite{Maki}. Besides this, the appearance of FFLO state also requires an extremely small numbers of disorders in the sample, as the impurity scattering  suppresses its formation \cite{11}.
Previous studies using bulk measurements such as the magnetostriction, ultrasound velocity and heat capacity have reported an upturn of the upper critical field and a transition inside the superconducting phase in heavy fermion superconductors\cite{Matsuda2007}, low-dimensional organic superconductors\cite{Wosnitza2018,17}, and transition metal dichalcogenide superconductors\cite{18,19}. Iron-based superconductors, known for their quasi-two-dimensional structure, strong Pauli paramagnetic effect, and multiband nature, are considered an ideal platform for investigating FFLO physics \cite{69}. Previous transport and specific heat measurements have suggested the potential presence of an FFLO state in FeSe above the upper critical field $B_{c2}$\cite{49,Kasahara2021}. However, the observed high-field superconducting states were found to be insensitive to disorders, which is inconsistent with the characteristics of the FFLO state\cite{48}. Notably, there is currently no microscopic experimental evidence supporting the existence of the FFLO state in iron-based superconductors.

NMR can serve as a powerful tool for investigating the FFLO state in bulk superconductors, directly detecting the superconducting spin smecticity, a property in analogy with liquid crystal, arising from the staggered distribution of normal and superconducting regions, resulting in line splitting or broadening in the spectrum\cite{76}. Furthermore, nodes in the order parameter form domain walls due to this staggered distribution, leading to a phase twist of $\pi$ in the superconducting phase and local modification of electronic density of states, creating Andreev bound states\cite{70,73}. The polarized quasiparticles spatially localized in the nodes can enhance the spin-lattice relaxation rate divided by temperature $1/T_1T$\cite{16}.
However, these two behaviors have not been simultaneously observed in one single superconducting material. For instance, previous studies on  CeCoIn$_5$\cite{Bianchi2002,Kumagai2006}, $\beta''$-(ET)$_2$SF$_5$CH$_2$CF$_2$SO$_3$\cite{Koutroulakis2016}, and Sr$_2$RuO$_4$\cite{4} only reported line splitting or broadening without any anomaly in $1/T_1T$, while no NMR line anomaly was reported in $\kappa$-(BEDT-TTF)$_2$Cu(NCS)$_2$\cite{16} and CeCu$_2$Si$_2$\cite{13}. Therefore, further experimental measurements remain essential to verify the existence of the FFLO state in general and especially for investigating multiband effects in the emergent materials.

KFe$_2$As$_2$ is particularly clean compared to other iron-based superconductors. The residual resistivity ratio of this compound can be even up to 3000\cite{Liu2013}.
Experiments have shown that the upper critical field $B_{c2}$ of KFe$_2$As$_2$ is decided by the Pauli paramagnetic effect with Maki parameter $\alpha_M \sim 3$\cite{27,28}.  Additionally, its anisotropic Fermi surface and superconducting order parameter together favor the formation of FFLO Cooper pairs\cite{30,31,56,57}. Previous magnetic torque and specific heat measurements have indeed observed the upturn of the upper critical field at low temperatures and high fields in KFe$_2$As$_2$ \cite{36}. However, these findings do not explicitly address the multiband effect on the FFLO state. Such effect can broaden the parameter space where FFLO states can form and affect boundaries between FFLO and HSC states\cite{7,33,34,35,65}. Therefore, direct microscopic experimental investigations are required for the putative state and for establishing a precise FFLO boundary in KFe$_2$As$_2$.  

In this work, we use $^{75}$As-NMR to address these issues. We found microscopic evidence for a FFLO state both from NMR spectra and 1/$T_1$$T$. We established a concrete boundary between the FFLO and homogeneous superconducting states, which is different from what is expected for a single-band superconductor. The single crystal samples were grown by the self-flux method that described in Ref.\cite{Wang2014,47}. The superconducting critical temperature $T_c$ of our sample is 3.8 K from the AC susceptibility measurement\cite{SM}, which is among the highest reported for this compound, indicating its high quality\cite{Liu2013}. Below $T$ = 1.5 K, measurements were conducted by using a $^3$He-$^4$He dilution refrigerator. To minimize the heating effects at low temperatures, we used a 30 $\mu$s long $\pi/2$ pulse. We adopt the same method as that of Pustogow $et$ $al$\cite{Pustogow2019} to test the heat-up effect induced by RF pulses in the superconducting state. We also confirmed that the spin-lattice relaxation rate 1/$T_1$ is nearly frequency independent. Subsequently, all $T_1$ was measured at the central position of the NMR line. Further details can be found in Fig. S3 and S12(a)\cite{SM}. The external magnetic field was applied parallel to the $ab$-plane with precise angle control achieved through an single axis rotator(see supplementary materials\cite{SM} Fig. S5). The NMR spectra were obtained by summing the fast Fourier transforms of the spin-echo signals, while the spin-lattice relaxation rate 1/$T_1$ was measured by the saturation-recovery method. Field values were calibrated using the resonance frequency of $^{63}$Cu in the copper coil ($^{63}$$K$ = 0.235\% at low temperatures).

	\begin{figure}
		\centering
		\includegraphics[width=0.9\textwidth]{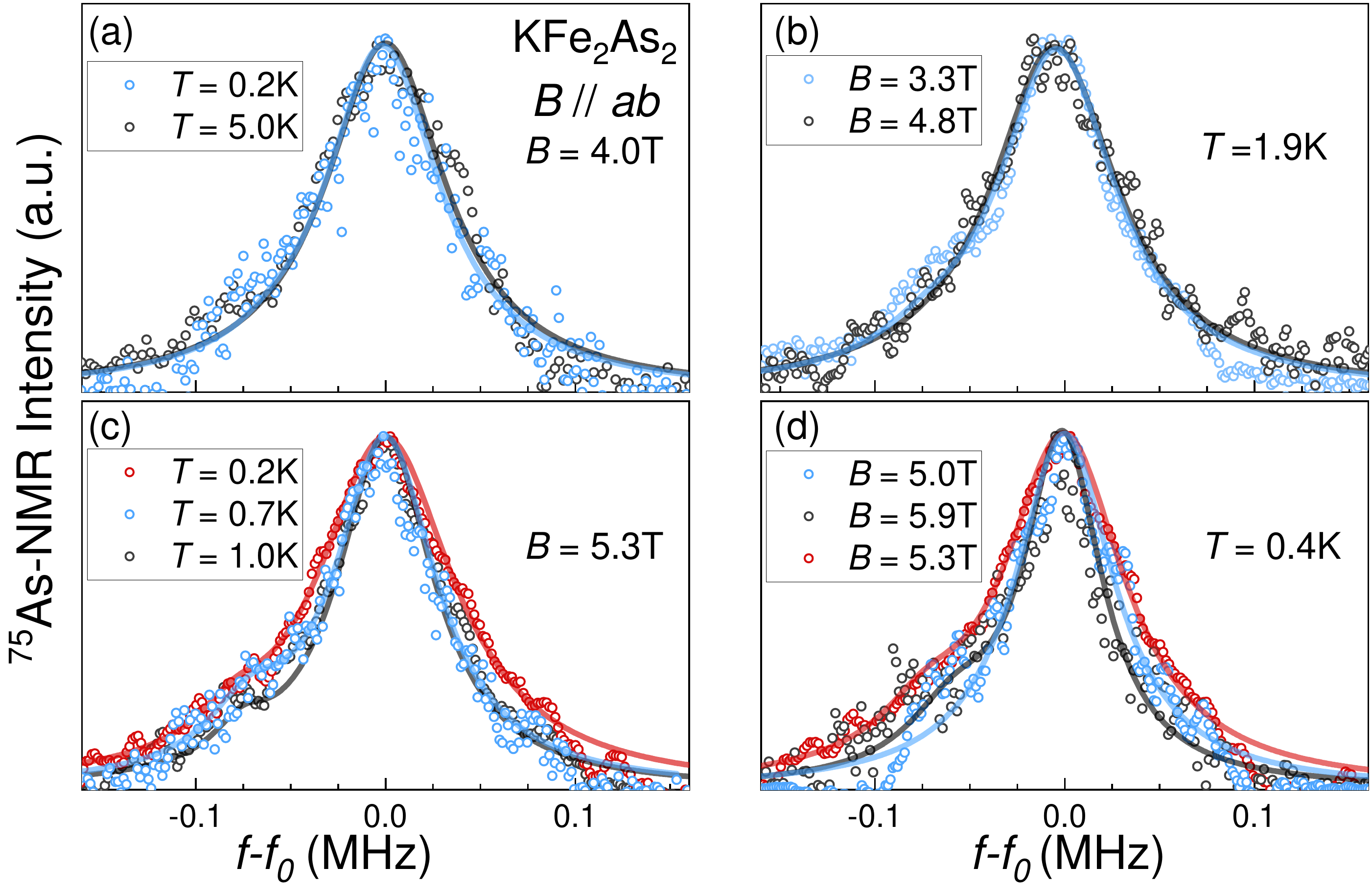}
		\caption{\label{fig:compare} (a) - (d) The central line of $^{75}$As-NMR spectra in the normal state(black color), the homogeneous superconducting state(blue color) and the FFLO state (red color) with the external magnetic field applied along the $ab$ plane (See Supplemental Materials for all spectra at various temperatures and fields\cite{SM}). $f_0$ ($\sim$ 24-45 MHz) is the frequency of the centre of the shown spectra.  Solid lines are guides to the eye. }
	\end{figure}

\begin{figure}
		\centering
		\includegraphics[width=0.9\textwidth]{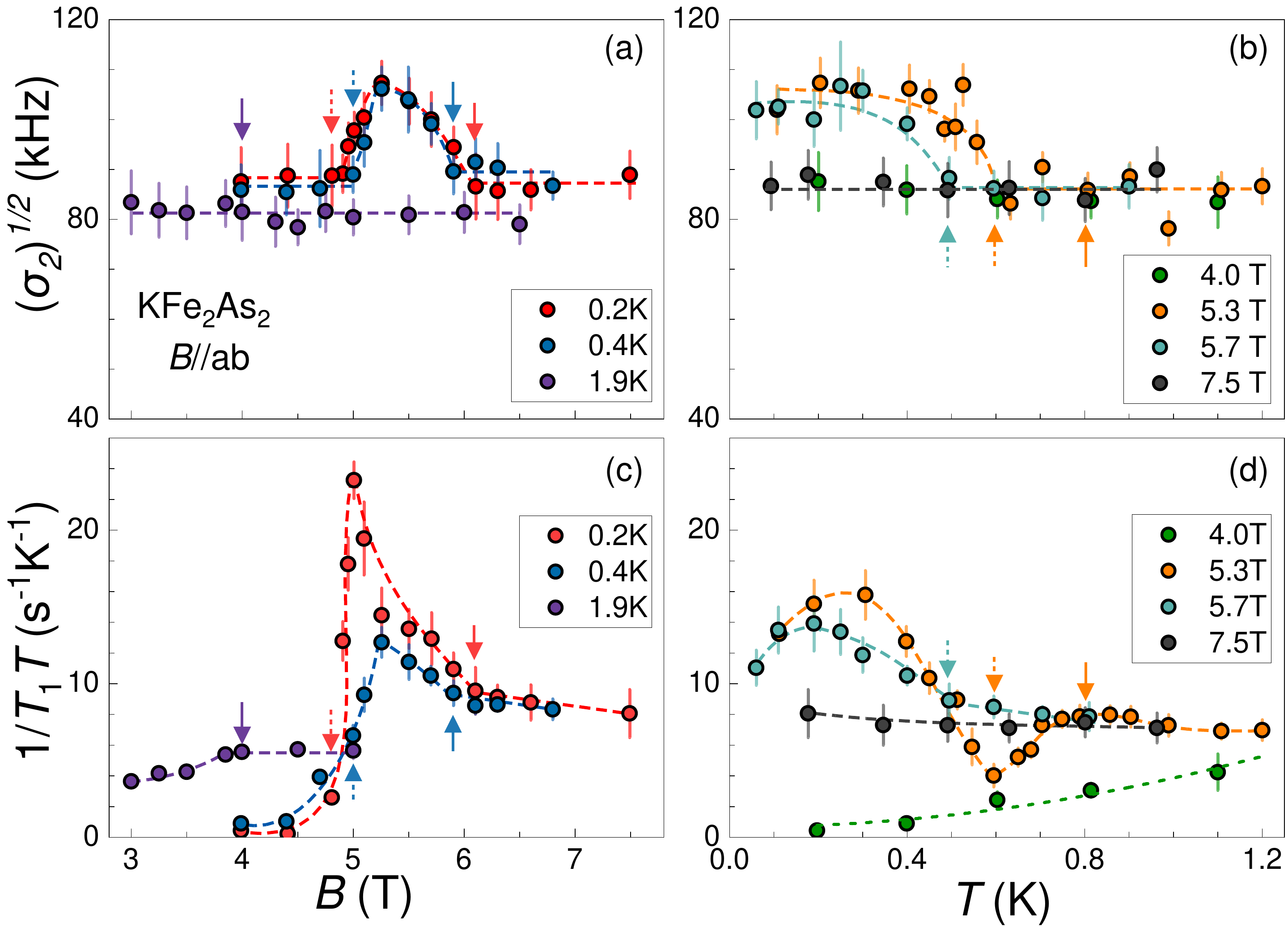}
		\caption{\label{fig:fwhmandt1}(a) and (c) are the field dependence of $(\sigma_2)^{1/2}$ of the spectrum and 1/$T_1$$T$ at various temperatures, respectively. (b) and (d) are the temperature dependence of the second moment of the spectra $(\sigma_2)^{1/2}$ and 1/$T_1$$T$ at various fields, respectively. The solid arrows denote a transition from the normal state to the HSC state. The dashed arrows identify a transition from the normal state or the HSC state to the FFLO state. Dashed lines are guides to the eye.}
	\end{figure}

	Figure \ref{fig:compare} presents the central transition line of the $\mathrm{^{75}As}$-NMR spectra at various temperature and magnetic fields. No spectral changes are observed with temperature down to 0.2 K at $B = 4$ T, nor with any magnetic fields at $T = 1.9$ K(Fig. \ref{fig:compare} (a) and (b)). However, at $B = 5.3$ T, line broadening appears at low temperatures below 0.6 K (Fig. \ref{fig:compare}(c)). At $T$ = 0.4 K, line broadening appears within a small field region 5 T $<$ $B$ $<$ 5.9 T(Fig. \ref{fig:compare}(d)).
To quantify this broadening, we calculate the square root of the second moment of the spectra $(\sigma_2)^{1/2}$\cite{SM}, which is a sensitive indicator of spin polarization \cite{4,13,40}.  The results of the calculation are presented in Fig. \ref{fig:fwhmandt1}(a) and \ref{fig:fwhmandt1}(b), revealing a clearly line broadening at low temperatures and high magnetic fields, at $T$ = 0.2 K with 4.8 T $< B <$ 6.1 T, $T$ = 0.4 K with 5 T $< B <$ 5.9 T, $T$ $<$ 0.6 K at $B$ = 5.3 T, and $T$ $<$ 0.5 K at $B$ = 5.7 T.
	In these regions, although no decrease in Knight shift was observed due to the broad central lines\cite{SM}, AC susceptibility measurements indicate that the NMR lines broadening indeed takes place within the superconducting state, even when the magnetic field exceeds the Pauli limit field $B_{\rm Pauli}$ = 4.8 T\cite{36}. When an external field is applied parallel to the superconducting layers of KFe$_2$As$_2$, the Josephson vortex state forms only between these layers and should not affect the linewidth, consistent with the temperature-independent behavior of $(\sigma_2)^{1/2}$ at $B$ = 4 T in the superconducting state (see Fig. \ref{fig:fwhmandt1}(b)).
	The enhancement of $(\sigma_2)^{1/2}$ of the spectrum occurs only within a magnetic field range of 4.8 T $< B <$ 6.1 T and below $T \approx$ 0.8 K, suggesting that the emergent line broadening is unrelated  to the vortex state. Instead, it arises from the superconducting spin smecticity, which is a characteristic feature of the FFLO state, as we elaborate below. In principle, the line splitting with a two-horn structure, as observed in Sr$_2$RuO$_4$\cite{4}, should be present(see Fig. S8(a))\cite{SM}. However, in KFe$_2$As$_2$, only line broadening is observed due to the relatively larger line width of the central line due to the second-order quadrupolar broadening (see Fig. S8(b) and (c))\cite{SM}.
We thus define the critical temperature $T^\ast$ or magnetic field $B^\ast$ of this possible FFLO state as the point at which the $(\sigma_2)^{1/2}$ of the spectrum begins to increase.
	
	
	
	\begin{figure}
		\centering
		\includegraphics[width=0.6\textwidth]{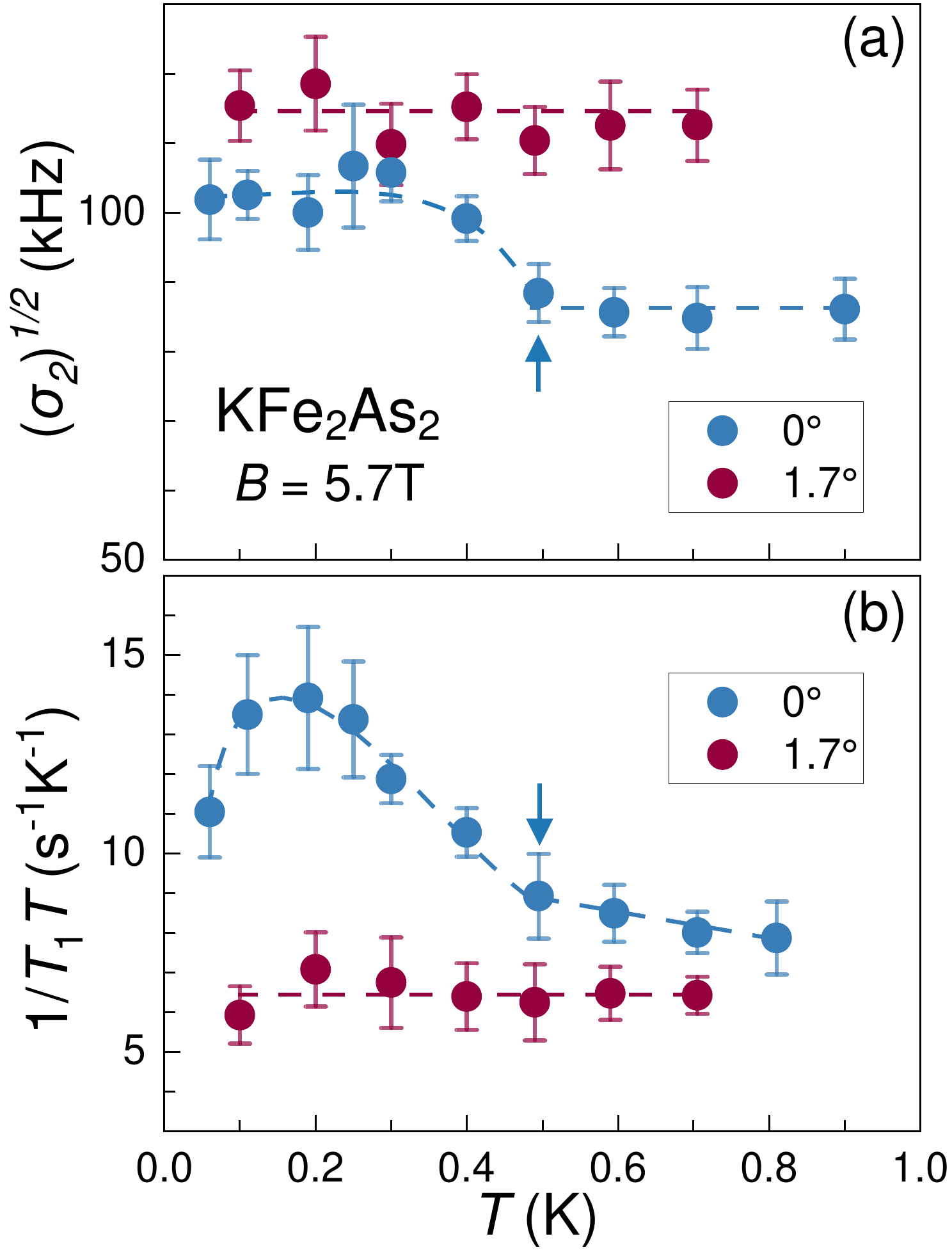}
		\caption{\label{fig:rotate} Comparison of the temperature dependence of $(\sigma_2)^{1/2}$ and 1/$T_1$$T$ between $\theta$ = 0$^\circ$(blue color) and 1.7$^\circ$(red color) at $B_0$ = 5.7 T in KFe$_2$As$_2$. The solid arrows represent a transition from the normal state to the  FFLO state. Dashed lines are guides to the eye.}
	\end{figure}

	To gain further insight into the inhomogeneous superconducting state, we measured the $1/T_1T$ as shown in Fig. \ref{fig:fwhmandt1}(c) and (d). At $B$ = 4 T, the temperature dependence of $1/T_1$ exhibits a rapid decrease with a $T^3$ variation at low temperatures, with no Hebel-Slichter coherence peak\cite{SM}, suggesting the unconventional nature of the superconductivity. The 1/$T_1$ $\propto$ $T^3$ behavior is in contrast with other iron-based superconductors\cite{OKA,Li2011}, but is consistent with the existence of line nodes in the gap function as previously observed by the angle-resolved photoemission spectroscopy (ARPES)\cite{30,31}.  A rapid reduction in $1/T_1T$ is also observed below the upper critical field $B_{c2}$ = 4 T at $T$ = 1.9 K (see Fig. \ref{fig:fwhmandt1}c).
	As the magnetic field increases, distinct peaks in $1/T_1T$ are observed at 5.3 T and 5.7 T, coinciding with the temperatures at which $(\sigma_2)^{1/2}$ of the NMR spectrum begins to increase. Additionally, at $T$ = 0.2 K and 0.4 K, peaks are also observed in the field variation of $1/T_1T$, at the magnetic fields where $(\sigma_2)^{1/2}$ increases (see Fig. \ref{fig:fwhmandt1}c). The correlation between the second moment $(\sigma_2)^{1/2}$ and the $1/T_1T$ indicates that the emergence of the superconducting spin smecticity and Andreev bound states occurs within the same temperature and high-field region. Our results provide compelling evidence for the existence of the FFLO state. Notably, at 5.3 T, there is a decrease in the $1/T_1T$ values in range  $T^\ast < T < T_c$, suggesting that the sample first transitions into the HSC state before evolving into the FFLO state as the temperature decreases.
In the FFLO state, one would also anticipate that 1/$T_1$$T$ could be frequency-dependent due to the spin smecticity. We indeed observed a frequency dependence of 1/$T_1$$T$ (see Fig. S12(b)), although rather weak. Nevertheless, the disparity of 1/$T_1$$T$ between the lowest and highest frequency points was within the error margin. The frequency-dependent behavior of 1/$T_1$$T$ could be smeared out in KFe$_2$As$_2$, as the NMR spectra merely exhibit line broadening, instead of the line splitting observed in Sr$_2$RuO$_4$\cite{4}. It would be interesting to conduct similar measurements on Sr$_2$RuO$_4$ in the FFLO state in the future.
At $B$ = 7.5 T, the $1/T_1T$ data show a slight upturn as the temperature decreases, which can be attributed to spin fluctuations\cite{Wang2016}.
	
	
	\begin{figure}
		\centering
		\includegraphics[width=0.9\textwidth]{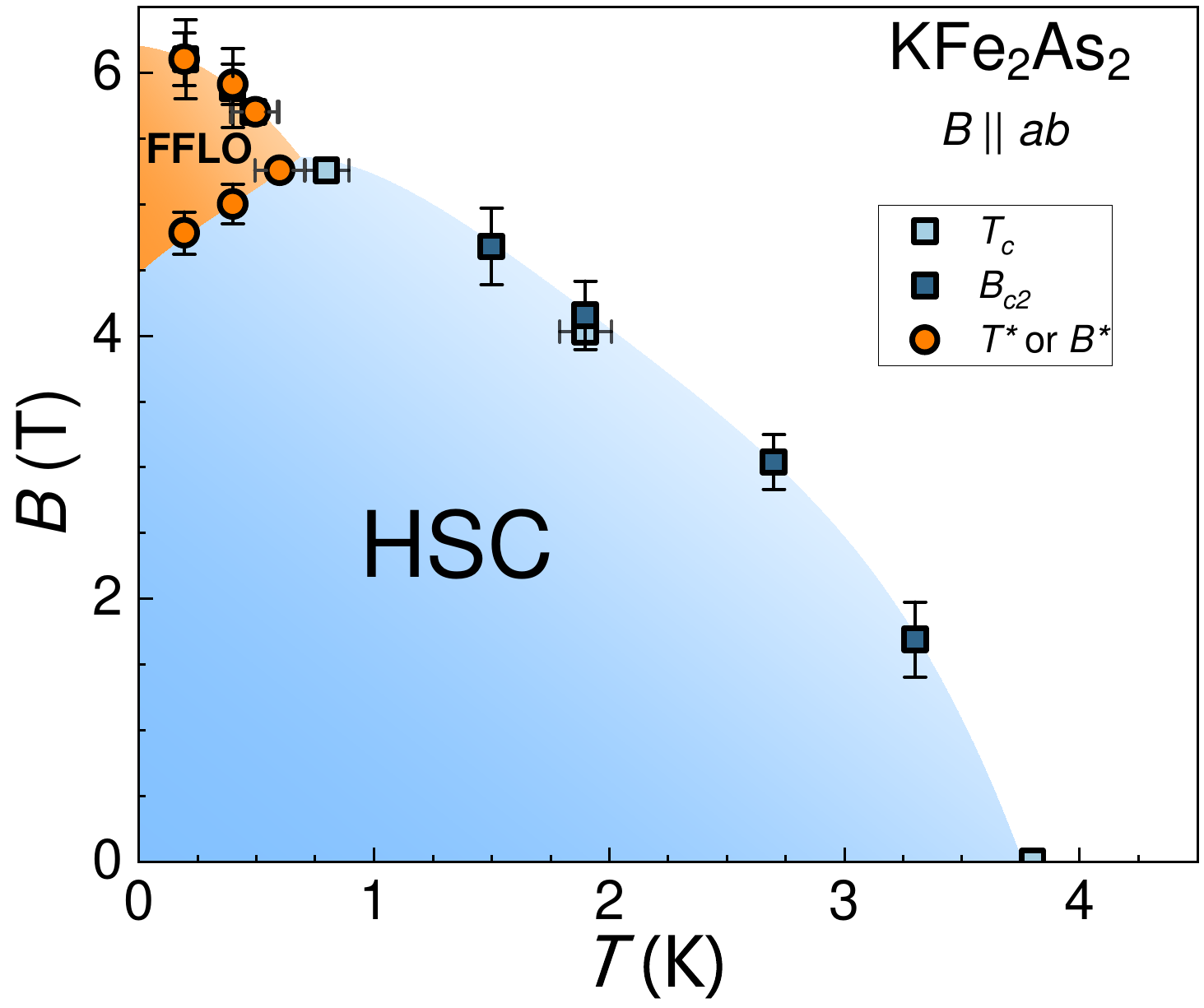}
		\caption{\label{fig:phase}The obtained phase diagram. The light blue and dark blue squares represent the superconducting critical temperature $T_c$ and the upper critical field $B_{c2}$ obtained from the AC susceptibility measurements\cite{SM}. The orange circle is the onset field $B^\ast$ and temperature $T^\ast$ determined based on the increase of $(\sigma_2)^{1/2}$ and 1/$T_1$$T$. HSC and FFLO denote the homogeneous superconducting state and the Fulde-Ferrell-Larkin-Ovchinnikov state, respectively.}
	\end{figure}

	For a FFLO state in quasi-two-dimensional superconductors, it should be sensitivity to the field orientation\cite{68}. The FFLO state can be significantly suppressed when the magnetic field deviates from being parallel to the superconducting layers, leading to the competition with the Abrikosov vortex state that forms within the superconducting layer. Upon rotating the sample by 1.7$^\circ$, we indeed found that $(\sigma_2)^{1/2}$ of the spectra becomes temperature-independent and the peak in $1/T_1T$ disappears, as shown in Fig. \ref{fig:rotate} (a) and (b), respectively.
These results show that the observed high-field state is indeed highly sensitive to the angle between the external field and the $ab$-plane, consistent with previous specific heat measurements\cite{36}.

We summarize our experimental findings in the phase diagram presented in Fig. \ref{fig:phase}.
Comments on two characteristic features found in this study are in order. Firstly, the transition field from HSC to FFLO state decreases as temperature decreases. The decreasing transition field for HSC to FFLO state with decreasing temperature suggests that the FFLO state may be more readily formed at lower temperatures. We ascribe this phenomenon to interband coupling in KFe$_2$As$_2$ as elaborated below. Because of the finite center-of-mass momenta of the Cooper pairs, the stability of the FFLO state strongly depends on the Fermi surface structure and the pairing anisotropy\cite{33}. Recent ARPES studies have demonstrated the presence of multiple Fermi surfaces with different anisotropies in KFe$_2$As$_2$\cite{30}. The $\alpha$ and $\gamma$ Fermi surface is nearly isotropy. The superconducting gap on the $\alpha$ Fermi surface is also nearly isotropic, and it is nearly zero on the $\gamma$ Fermi surface. In contrast, the $\beta$ and $\varepsilon$ Fermi surfaces display anisotropy, with line-nodal superconducting gap which can facilitate the FFLO formation. Additionally, ARPES measurements also suggested that the normal state of the $\varepsilon$ band features a very flat band, which can enhance the stabilization of the FFLO state through nesting\cite{Wosnitza2018,35}. Therefore the observed FFLO pairing in KFe$_2$As$_2$ can be mainly ascribed to the two anisotropic Fermi surfaces. Now, previous numerical calculations have shown that  the FFLO state can be  stabilized even in the lower field region due to interband coupling between an active band and a passive band where a gap is negligibly small\cite{33}. In the present case, the superconducting gap in the $\gamma$ band is much smaller that that in the other bands\cite{30}, so that the theory of Ref.\cite{33} can apply.
Namely, coupling between the $\gamma$ and other bands can result in a decrease in the lower critical field of the FFLO state at low temperatures and establishes an ascending boundary line between the FFLO state and HSC state as observed in the phase diagram of KFe$_2$As$_2$.

Secondly, we found a low critical temperature $T^\ast \approx$ 0.2$T_c$ of the FFLO state.
For the heavy fermion and organic superconductors, the $T^\ast /T_c$ ratio is larger than 0.4 in possible FFLO states\cite{Matsuda2007,Wosnitza2018,17}, where the Maki parameter is typically very large, implying that the orbital pair-breaking is usually negligible. For instance, in CeCu$_2$Si$_2$ and $\kappa$-(BEDT-TTF)$_2$Cu(NCS)$_2$, the Maki parameter $\alpha_M$ is $\sim 10$\cite{Kittaka2014} and $\sim 8$ \cite{Lortz2007}, in which $T^\ast /T_c$ ratio is even close to 0.6. It has been shown that orbital effects are detrimental to the formation of the FFLO state\cite{26}; the FFLO region will shrink due to the orbital effect, leading to a lower critical temperature. For KFe$_2$As$_2$, the Maki parameter is only $\sim$ 3, which is indeed much smaller than that of the heavy fermion and organic superconductors. Therefore, one possible explanation is that the orbital pair-breaking can not be ignored in KFe$_2$As$_2$. We note that an upturn in $B_{c2}(T)$ also appears at a relatively low temperature($T^\ast \approx 0.2 T_c$) in FeSe\cite{49}, where $\alpha_M$  is found to be similar to KFe$_2$As$_2$ as $\sim$ 5(2.5) for the electron(hole) pocket.  Therefore, it is possible that both compounds have a significant orbital pair-breaking, resulting in the occurrence of low critical temperature $T^\ast$. Additionally, considering that KFe$_2$As$_2$ is the multiband superconductor, the band structure may also be a factor influencing  the critical temperature $T^\ast$. The numerical calculations have shown that the electrons from the isotropic Fermi surface can strongly reduce the stability of the FFLO state and decreases its region in the phase diagram due to the quantum band geometric effects\cite{35}. Indeed, the $\alpha$ and $\gamma$ Fermi surfaces are nearly isotropy\cite{30} in KFe$_2$As$_2$, which might play a role.

We should also comment on the discrepancy from the previous transport measurements\cite{36}(see Fig. S9\cite{SM}). Some comments are in order. Previous transport measurements suggested that the FFLO state appears below $T^\ast \approx$ 1.5 K. However, our studies indicate that the FFLO state actually emerges at a lower temperature of $T^\ast \approx 0.8$ K even at $B\approx$ 5.3 T, coinciding with an upturn in the upper critical field. The presence of impurities in the sample is known to influence the appearance of the FFLO state, typically resulting in its occurrence at lower temperatures with higher impurity levels. However, our sample exhibits a higher $T_c$ than previously studied samples\cite{36,SM}, which is an indication of fewer impurities. It is unlikely that this difference can be attributed to impurity effects. We believe that the discrepancy arises with the large uncertainty in determined $T^\ast$ from the specific heat results\cite{36}. Finally, it is worth noting that while the $T_c$ of previously studied samples was found to be lower than ours\cite{36}, the temperature dependence of the upper critical magnetic field at $B$ $>$ 4 T is consistent with our findings\cite{SM}, suggesting that the observed FFLO state seems surprisingly robust against impurities. This behavior seems to contradict the presence of the FFLO state at high fields\cite{Wosnitza2018}. However, as already mentioned above that the isotropic superconducting gap on the $\alpha$ Fermi surface should contribute less to forming the FFLO state in KFe$_2$As$_2$, while the anisotropic gaps on $\beta$ and $\gamma$ Fermi surfaces might play a more significant role. Previous research has suggested the presence of interstitial Fe in KFe$_2$As$_2$\cite{Liu2013,Grinenko2013}, which serve as the magnetic impurity.
Therefore, it is plausible that although magnetic impurities  affect the isotropic superconducting gap and result in reduced $T_c$ at zero field, they may have minimal effect on the anisotropic superconducting gap and thus do not influence the presence of the FFLO state.  For the fully-opened gap with no sign-changing order parameter, the magnetic impurity is a pair breaker and breaks the time-reversal symmetry\cite{Anderson}. For the anisotropic superconducting gaps on $\beta$ and $\gamma$ Fermi surfaces, it is possible that the interband coupling makes them have more tolerance with magnetic impurities\cite{Li2009}. Further investigation is still required to study the impact of magnetic impurities on the FFLO state in KFe$_2$As$_2$ through the microscopic research on locally induced impurity states.

	In summary, we have preformed low-temperature and high-field $^{75}$As-NMR measurements on the multiband iron-based superconductor KFe$_2$As$_2$. A clear increase of the second moment of the spectrum, as well as a notable  peak in the $1/T_1T$ are observed in the same low-temperature and high-field region, implying the emergence of superconducting spin smecticity and Andreev bound states due to the staggered distribution of normal and superconducting regions, which provides compelling microscopic evidence for the existence of the FFLO state. The obtained phase diagram reveals two unique characteristics, a distinct boundary line between the FFLO and HSC states, and a low FFLO critical temperature $T^\ast \approx$ 0.2 $T_c$. These features can be attributed to the multiband effects. Our results offers new insights into the intrinsic properties and microscopic mechanisms underlying FFLO states.

\begin{acknowledgments}
We thank Hiroto Adachi, Takeshi Mizushima and Shigeru Kasahara for valuable discussions.
This work was supported by National Key Research and Development Projects of China (Grant No. 2023YFA1406103, No. 2022YFA1403402 and No. 2024YFA1611302), the National Natural Science Foundation of China (Grant No. 12374142 and No. 12304170), the Strategic Priority Research Program of the Chinese Academy of Sciences (Grant No.XDB33010100), Beijing National Laboratory
for Condensed Matter Physics (Grant No. 2024BNLCMPKF005) and CAS PIFI program (2024PG0003). J. Z. and Z. K. were supported by the Key Program of National Natural Science Foundation of China (Grant No. 12234006), National Key R$\&$D Program of China (Grant No. 2022YFA1403202), Innovation Program for Quantum Science and Technology (Grant No. 2024ZD0300103), and the Shanghai Municipal Science and Technology Major Project (Grant No. 2019SHZDZX01). This work was supported by the Synergetic Extreme Condition User Facility (SECUF, https://cstr.cn/31123.02.SECUF).
\end{acknowledgments}

\end{document}